\documentclass[twocolumn,preprintnumbers,amsmath,amssymb,a4paper,raggedbottom, balancelastpage,raggedfooter,floatfix]{revtex4}
\usepackage{graphicx}
\usepackage{bm}
\begin{document}
\renewcommand{\arraystretch}{1.2}

\title{Theoretical constraints on new generations with and without Quarks or
Neutrinos}

\author{Alexander Knochel}
\email{A.K.Knochel@thphys.uni-heidelberg.de}
\author{Christof Wetterich}%
 \email{C.Wetterich@thphys.uni-heidelberg.de}
\affiliation{%
Institut f\"ur Theoretische Physik, Ruprecht-Karls-Universit\"at Heidelberg,
Germany
}%

\begin{abstract}
We consider large classes of chiral extensions of the Standard Model, including
new quark generations that do not involve additional neutrinos as well as lepton
generations without quarks. An analysis of
renormalization flows of Yukawa and quartic scalar couplings reveals that
additional quarks are not compatible with a scenario of grand unification
without violating the strong bounds from direct and Higgs searches at colliders.
Constraints from direct searches, electroweak precision observables, and Higgs
physics, together with the assumption that additional new physics beyond the
extended chiral field content should enter significantly above the TeV scale,
allows us to make predictions for searches at the LHC. 
\end{abstract}


\maketitle
\section{Introduction}
The question whether there are additional chiral fermions beyond the known three
generations of quarks and leptons has so far found no conclusive answer. While
virtually all models of this type will be tested by the LHC experiments in the
near future, the presence of a fourth standard generation of quarks and leptons
is still a viable option.  However, constraints from flavour physics and the
absence of a fourth light neutrino suggest that an additional chiral generation,
if present, is not a mere heavy copy of the known three. We therefore also
consider alternative anomaly free representations which naturally accommodate
the absence of a fourth light neutrino and flavor mixing. 

Since the masses of such hypothetical particles must arise via the Higgs
mechanism, all such extensions of the Standard Model are tightly constrained due
to the tension between collider searches and precision observables on one side,
and the renormalization flow of couplings on the other side.  In this letter we
argue that all chiral extensions with additional quarks are incompatible with a
standard scenario of grand unification. Their existence would require new strong
interactions at scales below $10^7$ GeV. Whenever the scale of such new physics
is above a few TeV, the renormalization flow of couplings towards partial fixed
points makes such theories highly predictive for the possible values of the
masses of new quarks and the higgs particle, with important consequences for
searches at the LHC. We extend our discussion to quarkless generations of chiral
leptons which may still remain compatible with grand unification. 

\section{Anomaly free representations}

\begin{table}[htb]
\begin{tabular}[b]{|c|cl|}\hline
Doublets & Singlets & \\ \hline
 $(\alpha_1,\mathbf 2)_{A_1}$ & $(\overline \alpha_1,\mathbf 1)_{-A_1-\frac{1}{2}}$ & $(\overline
\alpha_1,\mathbf 1)_{-A_1+\frac{1}{2}}$ \\ 
$\displaystyle \vdots$ & $\displaystyle \vdots$ & \\
 $(\alpha_{n_D},\mathbf 2)_{A_{n_D}}$ & $(\overline \alpha_{n_D},\mathbf 1)_{-A_{n_D}-\frac{1}{2}}$ & $(\overline
\alpha_{n_D},\mathbf 1)_{-A_{n_D}+\frac{1}{2}}$ \\  \hline  
\multicolumn{3}{|c|}{Anomaly cancellation: $\displaystyle \sum_{i=1}^{n_D}
|\alpha_i| A_i=0$}\\ 
 \hline
\end{tabular}
\caption{The anomaly free hypercharge assignment for an arbitrary number of
isospin doublets which allow Yukawa couplings to the Higgs doublet.
Representations are given as $(SU(3),SU(2))_Y$ for left handed spinors. Electric
charges can be read off directly from the hypercharges of the $SU(2)$ singlets.
\label{tabsoln1}}
\end{table}

\begin{table}[htb]
\begin{tabular}[b]{|l||lll|}\hline
SM4${}_Y$ & $(\mathbf 3,\mathbf 2)_A$ & $(\overline{\mathbf 3} ,\mathbf
1)_{-A-\frac{1}{2}}$ & $(\overline{\mathbf 3},\mathbf 1)_{-A+\frac{1}{2}}$ \\ 
& $(\mathbf 1,\mathbf 2)_{-3A}$ &  $(\mathbf 1,\mathbf 1)_{3A-\frac{1}{2}}$ & $(\mathbf 1,\mathbf 1)_{3A+\frac{1}{2}}$\\ \hline
SM4Q${}_Y$& $(\mathbf 3,\mathbf 2)_A$ & $(\overline{\mathbf 3},\mathbf
1)_{-A-\frac{1}{2}}$ & $(\overline{\mathbf 3},\mathbf 1)_{-A+\frac{1}{2}}$ \\ 
&  $(\mathbf 3,\mathbf 2)_{-A}$ & $(\overline{\mathbf 3},\mathbf
1)_{+A-\frac{1}{2}}$ & $(\overline{\mathbf 3},\mathbf 1)_{+A+\frac{1}{2}}$  \\ \hline
SM4L${}_{\vec{Y}}$& $(\mathbf 1,\mathbf 2)_A$ & $(\mathbf 1,\mathbf 1)_{-A-\frac{1}{2}}$ & $(\mathbf 1,\mathbf 1)_{-A+\frac{1}{2}}$ \\ 
&  $(\mathbf 1,\mathbf 2)_{B}$ & $(\mathbf 1,\mathbf 1)_{-B-\frac{1}{2}}$ & $(\mathbf 1,\mathbf 1)_{-B+\frac{1}{2}}$  \\ 
&  $(\mathbf 1,\mathbf 2)_{C}$ & $(\mathbf 1,\mathbf 1)_{-C-\frac{1}{2}}$ & $(\mathbf 1,\mathbf 1)_{-C+\frac{1}{2}}$  \\ 
&  $(\mathbf 1,\mathbf 2)_{D}$ & $(\mathbf 1,\mathbf 1)_{-D-\frac{1}{2}}$ & $(\mathbf 1,\mathbf 1)_{-D+\frac{1}{2}}$  \\
& \multicolumn{3}{c|}{$A+B+C+D=0$}\\ \hline
SM4Q'${}_{\vec{Y}}$& $(\mathbf 3,\mathbf 2)_A$ & $(\overline{\mathbf 3},\mathbf 1)_{-A-\frac{1}{2}}$ &
$(\overline{\mathbf 3},\mathbf 1)_{-A+\frac{1}{2}}$ \\ 
&  $(\mathbf 3,\mathbf 2)_{B}$ & $(\overline{\mathbf 3},\mathbf 1
)_{-B-\frac{1}{2}}$ & $(\overline{\mathbf 3},\mathbf 1)_{-B+\frac{1}{2}}$  \\
&  $(\mathbf 3,\mathbf 2)_{C}$ & $(\overline{\mathbf 3},\mathbf
1)_{-C-\frac{1}{2}}$ & $(\overline{\mathbf 3},\mathbf 1)_{-C+\frac{1}{2}}$  \\ 
& \multicolumn{3}{c|}{$A+B+C=0$}\\ 
 \hline
\end{tabular}
\caption{Consistent hypercharge assignments for a family of quarks and leptons,
for a family of quarks, and for a family consisting of leptons only. Finally, as
an example with an odd number of doublets, a family of six quarks is shown which
follows the same pattern. Representations are given as $(SU(3),SU(2))_Y$ for
left handed spinors. Electric charges can be read off directly from the
hypercharges of the $SU(2)$ singlets.  \label{tabsoln}}
\end{table}
In this work we restrict ourselves to isospin doublets which become heavy via
the Higgs mechanism. As long as the field content is vectorlike with respect to the
$SU(3)_c$, there are four anomaly constraints:
\begin{align}
\label{anomalyconstraint}
SU(2)^2U(1):&&\sum_{\lefteqn{i {\mbox{($SU(2)$ doublets)}}}} N_{ci}Y_i =0\\
SU(3)^2U(1):&&\sum_{\lefteqn{i \mbox{($SU(3)$ triplets)}}} Y_i =0\\
U(1)^3:&&\sum_{\lefteqn{i \mbox{(all)}}}  N_{ci} Y_i^3=0 \\
G^2 U(1):&&\sum_{\lefteqn{i \mbox{(all)}}}  N_{ci} Y_i=0
\end{align}
where sums are over left handed Weyl spinors.  In addition, we would like to
allow Yukawa couplings which means for isospin doublets that we introduce
complete Dirac fermions in some $SU(3)_c$ representation which, in terms of
left-handed Weyl spinors, satisfy
\begin{align}\label{yukawaconstraint}Y(X_L)=-\frac{1}{2}-Y(u_R^c)=\frac{1}{2}-Y(d_R^c).\end{align}
For a number of $n_D$ independent doublets, this would leave us with up to $4 +2
n_D$ constraints for $3 n_D$ parameters and thus $n_D\geq 4$.  The Yukawa
conditions are fortunately not independent of the anomaly constraints. They
solve the mixed gravitational and $SU(3)^2 U(1)$ constraints for each isospin
doublet separately, and reduce the cubic constraint to the $SU(2)^2U(1)$
constraint, 
\begin{eqnarray}
2 Y(X_L)^3+Y(u_R^c)^3+Y(d_R^c)^3&=&-\frac{3}{2} Y(X_L) \\ 
2 Y(X_L)+Y(u_R^c)+Y(d_R^c)&=&0\,. 
\end{eqnarray}
This leaves us with a maximum number of $1+2 n_D$ constraints for $3 n_D$
parameters, and therefore $n_D-1$ free parameters. The general solution is shown
in Table \ref{tabsoln1}.

One may furthermore demand an even number of $SU(2)$ doublets in order to avoid
the global $SU(2)$ anomaly. We do not impose this restriction in view of a
possible embedding of the model in higher dimensions for which valid examples
with an odd number of four-dimensional chiral doublets are known. 
Finally, we only consider chiral representations which forbid gauge invariant
Dirac masses.  An extensive discussion of consistent chiral extensions of the SM
can be found in \cite{Foot:1988qx}.
In the following, we will study the interesting minimal solutions using color
triplets (``quarks'') and singlets (``leptons'') only. The solutions to the
anomaly and Yukawa constraints for these cases are shown in Table \ref{tabsoln}.
The two simplest nontrivial choices are a family of one quark doublet and one
lepton doublet as present in the Standard Model, or two quark doublets and no
color singlets. Note that, while the hypercharges of the two isospin doublet
quarks in SM4Q${}_Y$ differ only by a sign, the field content is chiral due to
the $SU(3)$ representations, which would not have been the case for a family of
two lepton doublets.  In SM4${}_Y$, a fourth Standard model family is obtained
for $Y=\frac16$. This already exhausts the possibilities if we restrict
ourselves to scenarios with one free parameter.  The next simplest case with an
even number of doublets consists of a family of four lepton doublets, which can
for example be obtained from SM4${}_Y$ by making the quarks color singlets while
retaining the multiplicity. This scenario, while having more free parameters,
generally allows a higher perturbativity cutoff due to weaker constraints from
collider searches, and naively improves the unification of gauge couplings. We
also revisit the family of three quark doublets which was found to arise from
compactifications of six-dimensional $SO(12)$ GUTs \cite{Wetterich:1985jx}. The
scenario considered there can be obtained from SM4Q' by setting
$A=B=\frac16~,~C=-\frac13$.

The electric charges of baryons and leptons are integer only for specific values
of $A$ for the examples in Table \ref{tabsoln}, namely $A=n+\frac16$ for SM4,
SM4Q and SM4Q'. For SM4Q' not all $A,B$ and $C$ can
obey this condition, and indeed exotic generations with half-integer charges for
baryons and leptons can be obtained from higher dimensional unification
\cite{Wetterich:1985jx}. For non-integer charged baryons or leptons the particle
with lowest mass must be stable. However, stable baryons with half-integer
charge have annihilated very efficiently in early cosmology due to their strong
interactions. Their present relic abundance, also in cosmic rays, is too low for
detection \cite{IW}. 

The examples SM4Q and SM4Q' involve new quark generations without additional
leptons. They are therefore compatible with the  LEP bound of three
light neutrinos belonging to weak doublets. However, they introduce a large
number of color-charged fields and are therefore strongly constrained by Higgs
searches. Generations with additional SM-like lepton doublets such as SM4
contain additional sterile neutrinos whose mass is not protected by the gauge
symmetries of the standard model. This may render such a scenario less
attractive.

\section{An RGE analysis}
\begin{figure}[htb]
\begin{center}
\includegraphics[width=8.5cm]{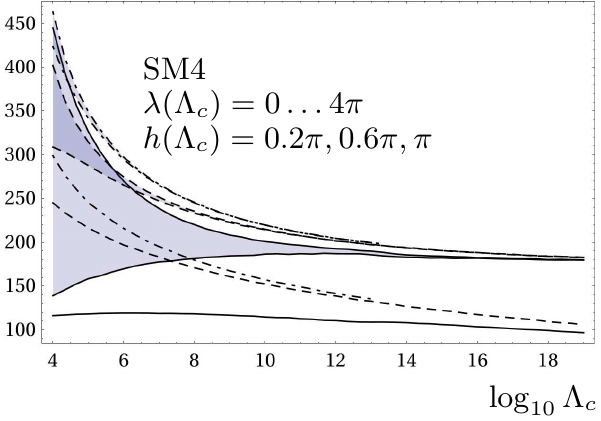}
\caption{\label{figfix_sm} An illustration of the coupled RG flow of the quartic
Higgs coupling and fourth generation quark Yukawa couplings in the SM4. The
(degenerate) fourth generation Yukawa couplings are $h(\Lambda_c)=0.2\pi
\mbox{(solid)}, 0.6\pi\mbox{(dashed)},\pi\mbox{(dot-dashed)}$.  For large fourth
generation Yukawa couplings, fourth generation masses saturate and $m_h$ (shaded
region) becomes independent of $\lambda$ as it approaches the partial fixed point. The
dot-dashed line ending at $\Lambda\sim10^{13}$ GeV indicates that the top quark
mass could not be accomodated for higher cutoffs. }
\end{center}
\end{figure}
\begin{figure}[htb]
\begin{center}
\includegraphics[width=8.5cm]{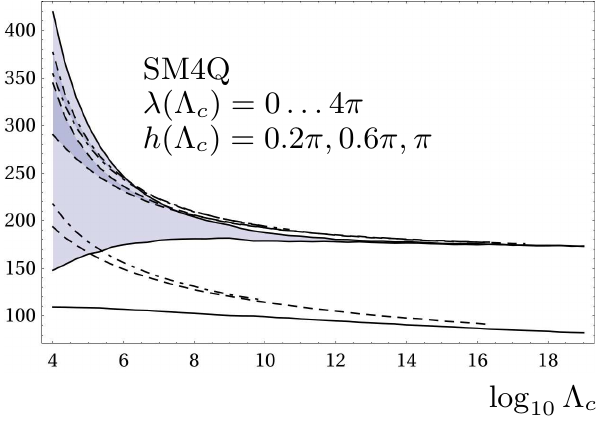}
\caption{\label{figfix_q4}
An illustration of the coupled RG flow of the quartic Higgs coupling and fourth
generation quark Yukawa couplings in the SM4Q. The (degenerate) fourth
generation Yukawa couplings are $h(\Lambda_c)=0.2\pi \mbox{(solid)},
0.6\pi\mbox{(dashed)},\pi\mbox{(dot-dashed)}$.  For large fourth generation
Yukawa couplings, fourth generation masses saturate and $m_h$ (shaded region)
becomes independent of $\lambda$ as it approaches the partial fixed point. The
dot-dashed/dashed lines ending at $\Lambda\sim 10^{10}/10^{16.5}$ GeV indicate
that the top quark mass could not be accomodated for higher cutoffs. 
}
\end{center}
\end{figure}
While the choice of Yukawa couplings and Higgs self coupling seems arbitrary from
a low-energy perspective, it turns out that the collective renormalization group
running drives the low-energy couplings close to a partial fixed point
\cite{CW2} if one chooses the input parameters at a scale $\Lambda_c >
\Lambda_{EW}$. For $\Lambda_c$ near some grand unified (GUT) scale, say
$10^{15}$ GeV or larger, one finds strong upper bounds for quark masses of
additional generations. They conflict with experimental bounds, such that models
with additional quark generations require new physics well below the GUT scale.
Even much lower $\Lambda_c$ significantly limits the accessible mass range for
the fourth generation fermions and the Higgs boson. This leads to predictive
scenarios which are subject to current LHC searches.  The situation is
illustrated in Figures \ref{figfix_sm} and \ref{figfix_q4}, where we show as a
function of $\Lambda_c$ the upper and lower bounds for the Higgs mass for
different (but universal) choices of the fourth generation quark Yukawa coupling
in the SM4 and SM4Q. The narrow range for the Higgs mass reflects the partial
fixed point in $\lambda/U^2_4$ discussed in \cite{CW2}. In these plots, it is
implied that the top quark mass is fixed at $m_t\sim 172$ GeV, which leads to
upper bounds on $\Lambda_c$ above which the top mass cannot be produced
perturbatively in the presence of the fourth generation Yukawa couplings.

The one-loop RGEs for the Higgs quartic and Yukawa coupling matrices for
arbitrary numbers of quarks and leptons are given by \cite{CW2} 
\begin{align}
\kappa^{-1}\beta_U&=\left(\frac{3}{2}(UU^\dagger - DD^\dagger) + \Sigma - 8 g_s^2 - ...\right) U \\
\kappa^{-1}\beta_D&=\left(\frac{3}{2}(DD^\dagger - UU^\dagger) + \Sigma - 8 g_s^2 - ...\right) D \\
\kappa^{-1}\beta_N&=\left(\frac{3}{2}(NN^\dagger - LL^\dagger)+\Sigma - ...\right) N \\
\kappa^{-1}\beta_L&=\left(\frac{3}{2}(LL^\dagger - NN^\dagger) +\Sigma -
...\right) L \\
\kappa^{-1}\beta_\lambda&=\left(12 \lambda^2 - 9 g_w^2 \lambda -\frac{9}{5} g^2
\lambda \right. \nonumber \\ &+ \left.  \frac{3}{4} \left(3 g_w^4 +\frac{6}{5} g_w^2 g^2 +\frac{9}{25} g^4\right)+ 4 \Sigma \lambda -4 X \right) 
\end{align}
where $\kappa^{-1}=16 \pi^2$ and we have defined
\begin{align}
X &= Tr[3 (U^\dagger U)^2+3 (D^\dagger D)^2 +(L^\dagger L)^2 + (N^\dagger
N)^2]\nonumber \\
\Sigma&=Tr[3 U^\dagger U+3 D^\dagger D +L^\dagger L + N^\dagger N].
\end{align}
The Yukawa couplings $U$ for the generalized up-type quarks are rectangular
$L_q\times R_u$ matrices, with $L_q$ the number of quark doublets, and $R_u$ the
number of up-type singlets. If the up-type quarks have different charges the
Yukawa matrix is block diagonal. The same holds for the other Yukawa couplings
$D$ for down-type quarks, $L$ for charged leptons and $N$ for neutrino type
particles. (Note that neutrino type particles are neutral only for a particular
value of the hypercharge in Table \ref{tabsoln}). Our convention assumes a tree
level Higgs potential of the form $\mu^2 \phi^\dagger \phi +\lambda/2
(\phi^\dagger \phi)^2$. In our numerical analysis, we use the two-loop results
from \cite{Luo:2002ey},\cite{Luo:2002ti}, neglecting the contributions from
electroweak gauge couplings.
\begin{figure}[htb]
\begin{center}
\includegraphics[width=8cm]{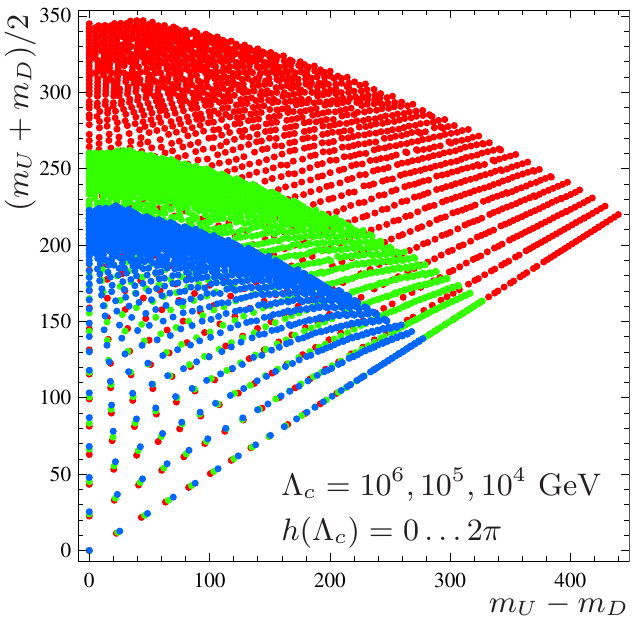}
\end{center}
\caption{The accessible mass range of fourth generation quarks in the SM4 for
high scales $\Lambda_c=1000$\mbox{(blue)},$100$\mbox{(green)},$10$\mbox{(red)} TeV. The leptons are fixed at $140$ GeV and
$60$ GeV. \\ \label{sm4quarkmasses}}
\end{figure}
\begin{figure}[htb]
\begin{center}
\includegraphics[width=8cm]{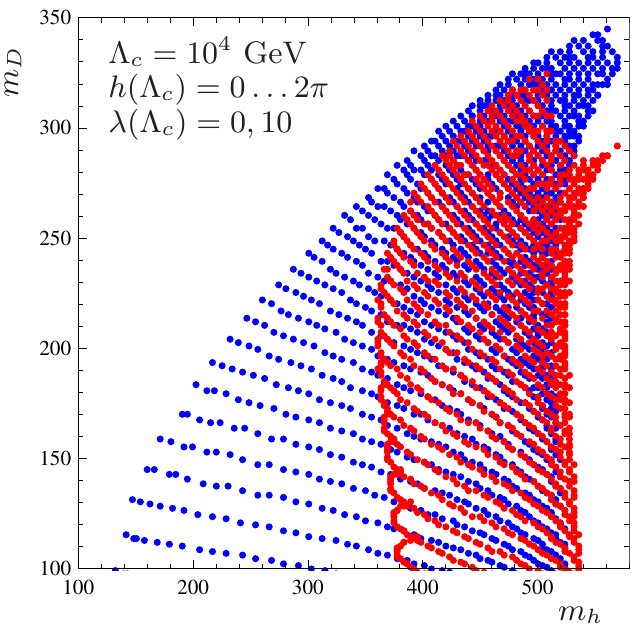}
\end{center}
\caption{The accessible mass range of the lightest fourth generation quark and
the Higgs boson for $\lambda(\Lambda_c)=0$ (blue) and $\lambda(\Lambda_c)=10$
(red) in the SM4.  \label{sm4higgscomp}}
\end{figure}
\begin{figure}[htb]
\begin{center}
\includegraphics[width=8cm]{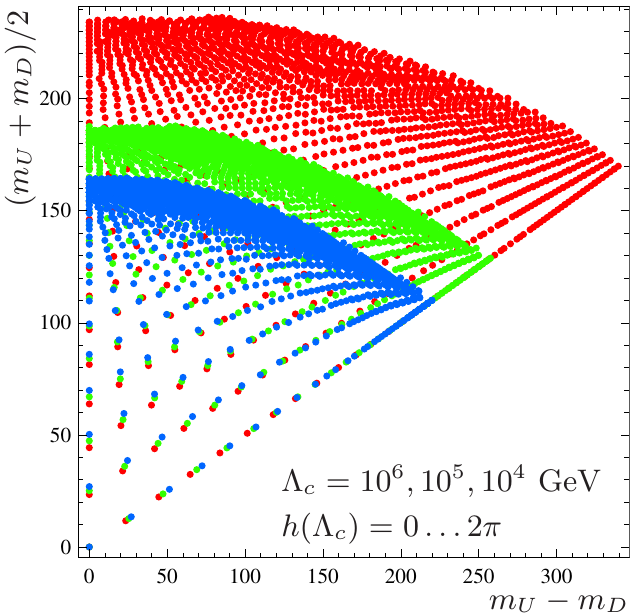}
\end{center}
\caption{The accessible mass range of fourth generation quarks in the SM4Q for
high scales $\Lambda_c=1000$\mbox{(blue)},$100$\mbox{(green)},$10$\mbox{(red)}
TeV. The four quark masses are $(m_U,m_D)$ and $(m_D,m_U)$. \label{q4quarkmasses}}
\end{figure}
\begin{figure}[htb]
\begin{center}
\includegraphics[width=8cm]{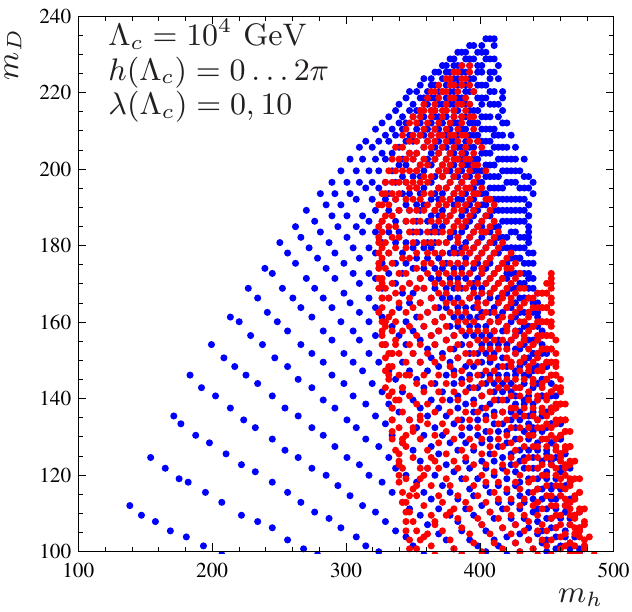}
\end{center}
\caption{The accessible mass range of the lightest fourth generation quark and
the Higgs boson for $\lambda(\Lambda_c)=0$ (blue) and $\lambda(\Lambda_c)=10$
(red) in the SM4Q. \label{q4higgscomp}}
\end{figure}
\begin{figure}[htb]
\begin{center}
\includegraphics[width=8cm]{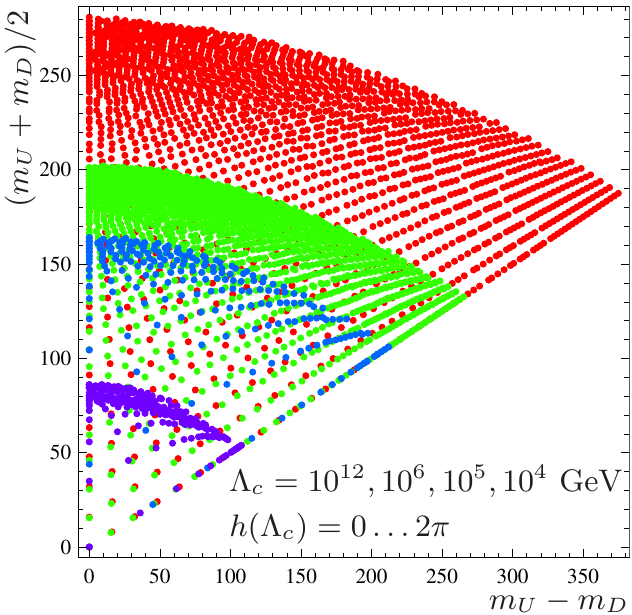}
\end{center}
\caption{The accessible mass range of fourth generation leptons in the SM4L for
high scales $\Lambda_c=10^9$ \mbox{(violet)},
$1000$\mbox{(blue)},$100$\mbox{(green)},$10$\mbox{(red)} TeV. The masses of the four lepton
doublets are chosen to be $3\times (m_U,m_D)$ and $(m_D,m_U)$.
\label{l4quarkmasses}} 
\end{figure}
\begin{figure}[htb]
\begin{center}
\includegraphics[width=8cm]{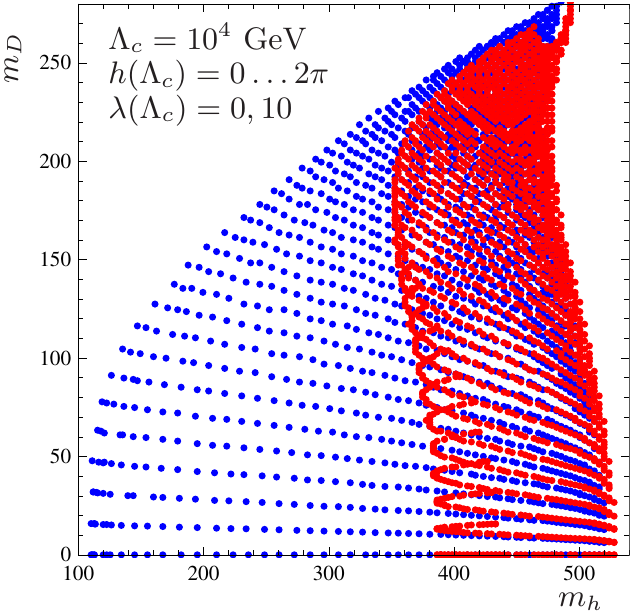}
\end{center}
\caption{
The accessible mass range of the lightest fourth generation quark and
the Higgs boson for $\lambda(\Lambda_c)=0$ (blue) and $\lambda(\Lambda_c)=10$
(red) in the SM4L.
\label{l4higgscomp}}
\end{figure}

We can immediately see that large Yukawa couplings for any field will tend to
drive the Yukawa couplings collectively towards smaller values due to the
universal positive contribution from $\Sigma$, while the running of the quartic
coupling receives contributions in both directions from $X$ and $\Sigma$. It is
well known that for $\Lambda_c<M_{Pl}$, a relatively large range of physical
Higgs boson masses can be reached in the SM, while additional chiral particle
content contributing to $X$ and $\Sigma$ will dominate the running of $\lambda$
and lead to a more precise prediction of $m_h$ \cite{Wetterich:1985jx}. This
behavior is illustrated in Figures \ref{figfix_sm} and \ref{figfix_q4} for the
SM4 and SM4Q.

One can now choose a scale $\Lambda_c$ at which one defines the input parameters
(Yukawa and quartic couplings), and evolve them down to the electroweak scale. A
scan over the high scale parameters then reveals the accessible mass range of
the Higgs and fourth generation particles as a function of $\Lambda_c$. In this
analysis, we use the matching condition $M(\mu_0)=\mu_0$. The upper bound for
the quark mass reflects ``triviality'' of the running Yukawa couplings. It can
be obtained by starting formally with infinite Yukawa couplings $U_4$ at
$\Lambda_c$ - all smaller initial values will lead to smaller quark masses.  For
$\Lambda_c$ of the order of a GUT scale $10^{15}$ GeV and $m_t=172$ GeV, the
$t'$ and $b'$ quark masses have an upper bound $m_{t'}=m_{b'} \leq 135$ GeV. If
the bound is saturated, the model predicts a Higgs mass $m_H\sim192$ GeV. These
exotic quark masses are far outside present experimental bounds.  Bounds for
SM4Q~($m_{t'_i}=m_{b'_i}\leq 100$ GeV) or SM4Q'~($m_{t'_i}=m_{b'_i}\leq 82$ GeV)
are even stronger. We conclude that additional quark generations are not
compatible with grand unification. 

In this context we denote by ``grand unification'' a class of models with gauge
group $SU(3)\times SU(2)\times U(1)$ (or slight extensions as additional $U(1)$
factors) that are valid up to a high scale (say $\Lambda_c>10^{12}$ GeV) without
invoking new strong interactions involving so far undetected particles at
intermediate scales. The content of chiral fermions is left arbitrary.
(Additional vectorlike fermions or an extended Higgs sector will not modify
substantially the upper bound for $m_{q_4}$.) The only way to escape the upper
bound for $m_{q_4}$ in this setting would be a replacement of the triviality of
Yukawa couplings by a fixed point behavior for large values of the Yukawa
couplings \cite{CW2}, \cite{FP1}, \cite{GJW}, \cite{FP2}. So far, lattice
\cite{LL1}-\cite{LL4} or functional renormalization group studies \cite{CW2},
\cite{FP1}, \cite{GJW}, \cite{FP2} have not found such a fixed point in the
``grand unification setting''. (A possible exception could be models with strong
four-fermion interactions \cite{GJW}.)

Abandoning the GUT scenario new strong interactions would have to occur at
rather low scales $\Lambda_c$ in a range below $10^7$ GeV. In the following we
investigate this ``strong interaction scenario''. We show that for
$\Lambda_c>10$ TeV the heavy particle masses are already strongly constrained.
The results for the fourth generation quark masses in the SM4 are shown in
Figure \ref{sm4quarkmasses}. We emphasize the strong clustering of points near
the upper bound.  The corresponding results for the lepton-less scenario SM4Q
and the quark-less scenario SM4L are shown in Figures \ref{q4quarkmasses} and
\ref{l4quarkmasses}.  The dependence on the value  of the quartic coupling at
the high scale is strongly reduced for large fourth generation masses. This is
illustrated in Figures \ref{sm4higgscomp},\ref{q4higgscomp} and
\ref{l4higgscomp}.

We have considered the quark-less scenario SM4L at a very high cutoff scale. The
motivation for this is that the extension of the SM by color-neutral fields can
improve the precision of gauge unification, and the scale of unification without
additional color-charged fields (at 1-loop) is around $10^{12}$ GeV.
Furthermore, the masses in this scenario are least constrained by direct
searches. We find that the the upper bounds on the exotic lepton masses (which
can not all be made neutral) are very close to or below the exclusion bounds
from collider searches.  

\section{Electroweak Precision Tests}
\begin{figure}[htb]
\begin{center}
\includegraphics[width=8cm]{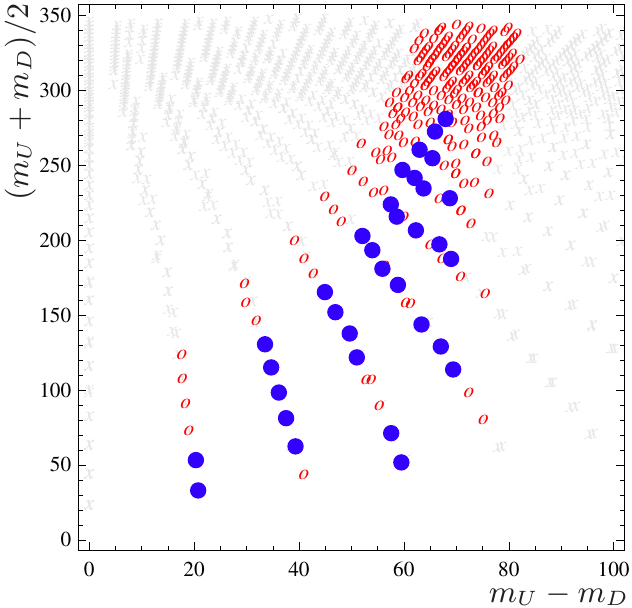}
\end{center}
\caption{
The points in the SM4 ($Y=\frac16$, $\Lambda_c=10$ TeV,
$\lambda(\Lambda_c)=0,10$) which are not excluded by electroweak precision tests at 68\%CL/95\%CL are shown as
blue dots/red circles.\label{sm4stu}}
\end{figure}

\begin{figure}[htb]
\begin{center}
\includegraphics[width=8cm]{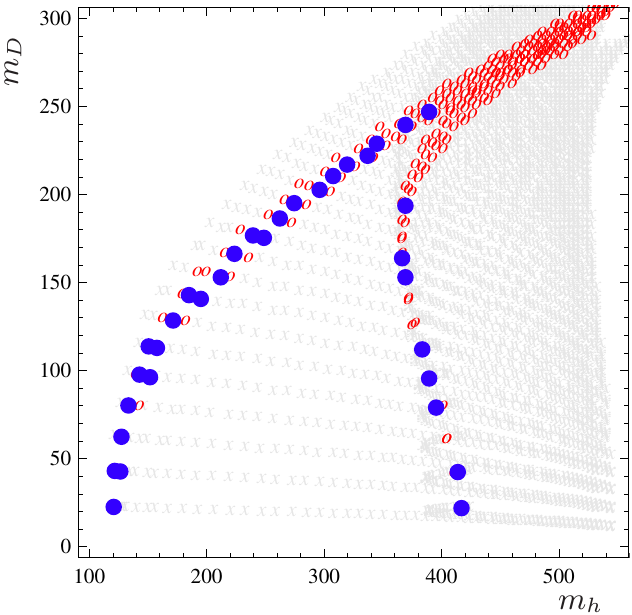}
\end{center}
\caption{
The points in the SM4 ($Y=\frac16$, $\Lambda_c=10$ TeV,
$\lambda(\Lambda_c)=0,10$) which are not excluded by electroweak precision tests at 68\%CL/95\%CL are shown as
blue dots/red circles.\label{sm4stuhiggs}}
\end{figure}

\begin{figure}[htb]
\begin{center}
\includegraphics[width=8cm]{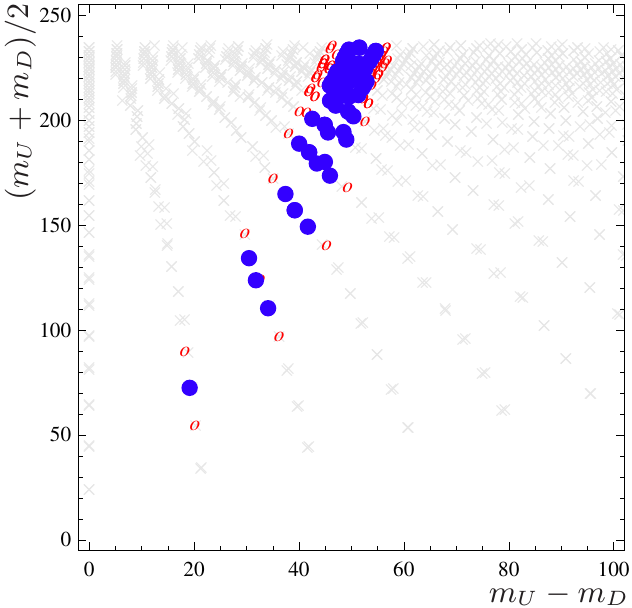}
\end{center}
\caption{
The points in the SM4Q ($Y=1$, $\Lambda_c=10$ TeV, $\lambda(\Lambda_c)=0,10$)
which are not excluded by electroweak precision tests at 68\%CL/95\%CL are shown as
blue dots/red circles. \label{q4stu}}
\end{figure}

\begin{figure}[htb]
\begin{center}
\includegraphics[width=8cm]{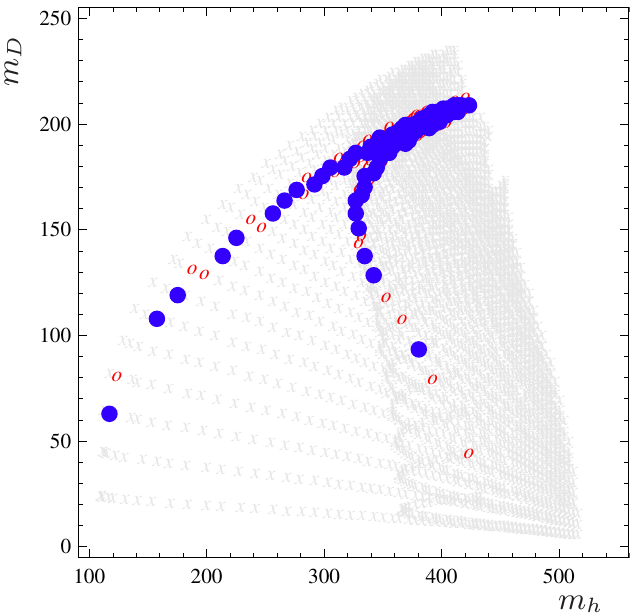}
\end{center}
\caption{
The points in the SM4Q ($Y=1$, $\Lambda_c=10$ TeV, $\lambda(\Lambda_c)=0,10$)
which are not excluded by electroweak precision tests at 68\%CL/95\%CL are shown as
blue dots/red circles. \label{q4stuhiggs}}
\end{figure}

\begin{figure}[htb]
\begin{center}
\includegraphics[width=8cm]{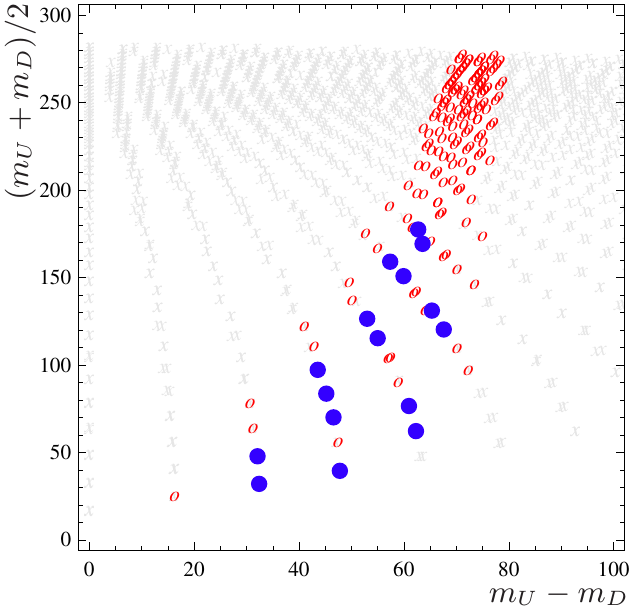}
\end{center}
\caption{
The points in the SM4L ($Y=\frac16$, $\Lambda_c=10$ TeV,
$\lambda(\Lambda_c)=0,10$) which are not excluded by electroweak precision tests at 68\%CL/95\%CL are shown as
blue dots/red circles.\label{l4stu1}}
\end{figure}

\begin{figure}[htb]
\begin{center}
\includegraphics[width=8cm]{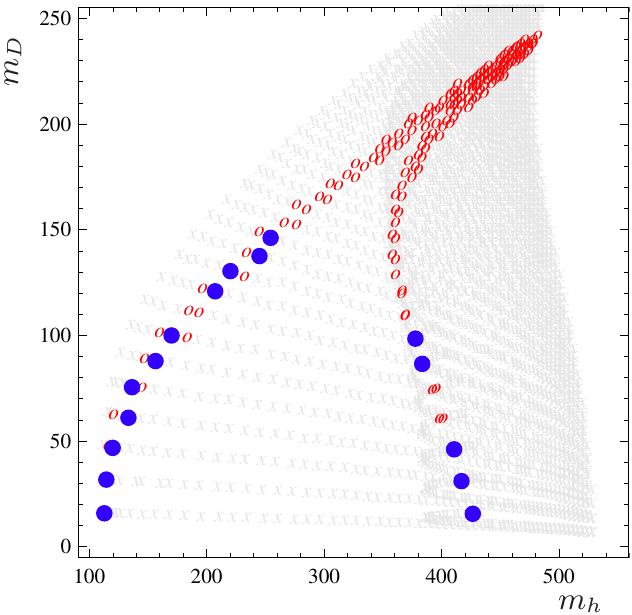}
\end{center}
\caption{
The points in the SM4L ($Y=\frac16$, $\Lambda_c=10$ TeV,
$\lambda(\Lambda_c)=0,10$) which are not excluded by electroweak precision tests at 68\%CL/95\%CL are shown as
blue dots/red circles.\label{l4stu1higgs}}
\end{figure}

\begin{figure}[htb]
\begin{center}
\includegraphics[width=8cm]{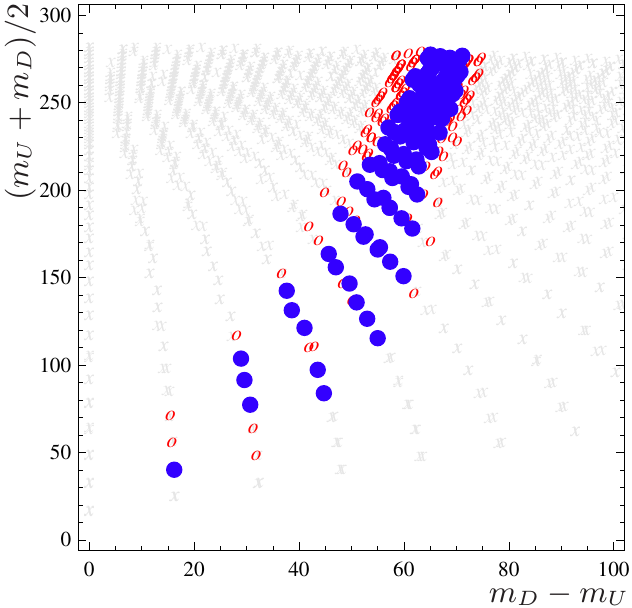}
\end{center}
\caption{
The points in the SM4L ($Y=-\frac12$, $\Lambda_c=10$ TeV,
$\lambda(\Lambda_c)=0,10$) which are not excluded by electroweak precision tests at 68\%CL/95\%CL are shown as
blue dots/red circles.\label{l4stu2}}
\end{figure}

\begin{figure}[htb]
\begin{center}
\includegraphics[width=8cm]{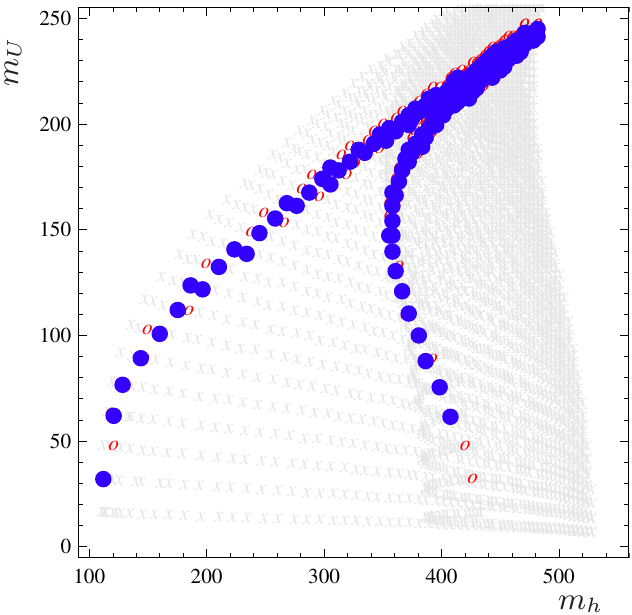}
\end{center}
\caption{
The points in the SM4L ($Y=-\frac12$, $\Lambda_c=10$ TeV,
$\lambda(\Lambda_c)=0,10$) which are not excluded by electroweak precision tests at 68\%CL/95\%CL are shown as
blue dots/red circles.\label{l4stu2higgs}}
\end{figure}

\begin{figure}[bt]
\begin{center}
\includegraphics[width=8cm]{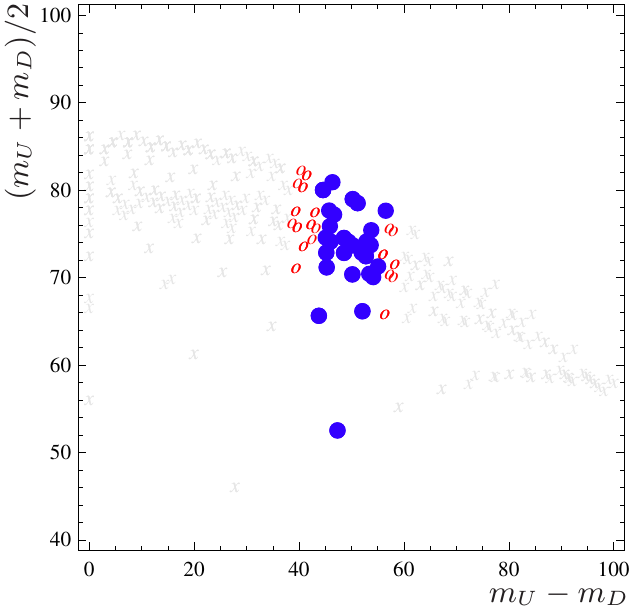}
\end{center}
\caption{
The points in the SM4L ($Y=\frac16$, $\Lambda_c=10^{12}$ GeV,
$\lambda(\Lambda_c)=0,10$) which are not excluded by electroweak
precision tests at 68CL/95CL are shown as
blue dots/red circles.
\label{l4stu3}}
\end{figure}

\begin{figure}[bt]
\begin{center}
\includegraphics[width=8cm]{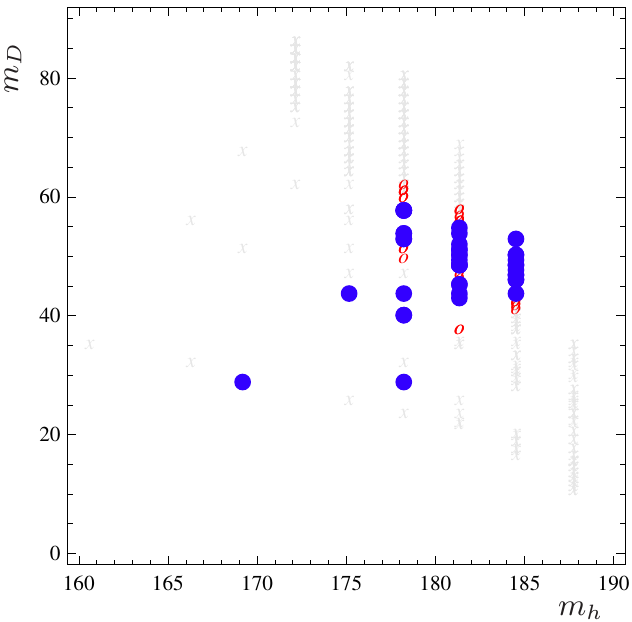}
\end{center}
\caption{
The points in the SM4L ($Y=\frac16$, $\Lambda_c=10^{12}$ GeV,
$\lambda(\Lambda_c)=0,10$) which are not excluded by electroweak
precision tests at 68CL/95CL are shown as
blue dots/red circles.
\label{l4stu3higgs}}
\end{figure}

We use the approximate expressions for the contributions of heavy isospin
doublets to the Peskin-Takeuchi $S$ and $T$ parameters
\cite{Peskin:1991sw},\cite{Kniehl:1992ez}, 
\begin{eqnarray}
\Delta T&=&\frac{N_c}{16 \pi s_w^2 m_W^2}\left(m_U^2 + m_D^2-\frac{m_U^2
m_D^2}{m_U^2-m_D^2}\log\frac{m_U^2}{m_D^2}\right) \nonumber \\
\Delta S&=&\frac{N_c}{6 \pi}\left(1-2 Y \log\frac{m_U^2}{m_D^2} \right)
\end{eqnarray}
which have been considered in a similar analysis by \cite{Kribs:2007nz} and agree with
the full electroweak calculation sufficiently for our purposes. $\Delta T$,
which measures the violation of custodial symmetry by the splitting of isospin
doublets, is positive semidefinite, while $\Delta S$ can be made small or
negative by the splitting.  This works best if $|Y|$ of the doublets is large.
The contributions to $S$ and $T$ from physics beyond the SM are tightly
constrained by experiments \cite{Nakamura:2010zzi},\cite{Baak:2009zz}. In the $\Lambda\geq 10$ TeV
scenarios with heavy fourth generation Quarks, we generically find Higgs masses
$m_H > 200\dots 300$ GeV. Such a heavy Higgs boson contributes to $S$ and $T$,
shifting the preferred range for $\Delta S$ and $\Delta T$ to smaller/larger
values respectively. In order to combine our results from the RGE analysis with
the electroweak constraints in a meaningful way, we check for each point of the
scan whether $\Delta S$ and $\Delta T$ lie withing the $68\%CL$ or the $95\%CL$ ellipse given
by \cite{Baak:2009zz},\cite{Nakamura:2010zzi}. First we consider the SM4 with standard hypercharge
assignments $Y(Q)=\frac16$. The result is shown in Figures
\ref{sm4stu},\ref{sm4stuhiggs}. The
lepton and neutrino masses which are not shown are fixed at $140$ GeV and $60$
GeV respectively.  The SM4Q does not pass electroweak precision tests for the
SM-like hypercharge assignment $Y(Q)=\frac16$ due to the large field content.
However, for $Y(Q)=1$, the cancellation of $\Delta S$ is more effective and a
range of masses becomes allowed by precision tests (Figures
\ref{q4stu},\ref{q4stuhiggs}). The
SM4L for $Y(L_{1\dots 3})=\frac16$ and $Y(L_4)=-\frac12$ yields results similar
to the SM4 case due to the identical field content and hypercharges (Figures
 \ref{l4stu1},\ref{l4stu1higgs}). As an interesting variation, we consider the case $Y(L_{1\dots
3})=-\frac12$ and $Y(L_4)=\frac32$ where three doublets are SM-like with neutral
neutrinos, while the fourth doublet has electrical charges $1$ and $2$. In this
case, having light exotic neutrinos gives small corrections to $S$ and is
advantageous in order to evade direct searches. Consequently, a relatively large
range of parameter space is allowed by electroweak precision tests (Figures
\ref{l4stu2},\ref{l4stu2higgs}). We can raise the scale to $\Lambda_c=10^{12}$ GeV and still find
a range of viable parameter points for the SM-like charge assignment (Figures
\ref{l4stu3},\ref{l4stu3higgs}).
\section{Higgs production}
The contributions of chiral matter to the effective $Hgg$, and $H\gamma\gamma$
operators do not decouple for $m\gg m_H$, which makes Higgs production and decay
an important probe of models with chiral particle content beyond the SM. In
particular at the LHC, the production of Higgs bosons is dominated by
$gg\rightarrow H$ for the entire accessible mass range, which results in a
production rate roughly proportional to $N_f^2$ where $N_f$ is the number of
heavy $SU(3)$ triplets. The coupling to photons receives contributions from $W$
boson loops, which interfere destructively with heavy charged fermionic matter,
which generically leads to a reduction of $\Gamma(H\rightarrow \gamma \gamma)$
except in extreme cases with many or multiply charged particles.  The partial
widths for $H \to \gamma \gamma$ and $H \to gg$ can be written
as~\cite{Ellis:1975ap},\cite{Kribs:2007nz} 
\begin{alignat}{5}
 \Gamma_{H \to \gamma \gamma} &=  \frac{G_\mu \alpha^2 m_H^3}{128
  \sqrt{2} \pi^3} \left| \sum_{f} N_c Q_f^2 A_f(\tau_f) + A_W(\tau_W)
  \right|^2
                \notag \\
  \Gamma_{H \to g g} &=  \frac{G_\mu \alpha_s^2 m_H^3}{36
  \sqrt{2} \pi^3} \left| \frac{3}{4} \sum_{f}  A_f(\tau_f) \right|^2 \; .
\label{eq:hgg}
\end{alignat}
where the form factors $A_f$ and $A_W$ for $s=\frac12$ and $s=1$ fields are
given by 
\begin{alignat}{5}
 A_f(\tau) &= 2 \left[  \tau
                      + (\tau - 1) f(\tau)
                \right] \, \tau^{-2} \notag \\
 A_W(\tau) &= - \left[  2 \tau^2
                      + 3 \tau + 3(2 \tau -1) f(\tau)
                \right] \, \tau^{-2}
\end{alignat}
with $\tau_i = m_H^2/4 m_i^2$, $(i=f,W)$ and
\begin{alignat}{5}
f(\tau)= \left\{
                 \begin{array}{cc}
                 \mathrm{arcsin}^2 \sqrt{\tau} & \tau \le 1 \\
                -\frac{\displaystyle 1}{\displaystyle 4}
                             \left[ \ln \frac{\displaystyle
                                               1+\sqrt{1-\tau^{-1}}}
                                              {\displaystyle
                                               1-\sqrt{1-\tau^{-1}}}
                                    -i \pi
                             \right]^2        & \tau > 1
\end{array}
\right. 
\end{alignat}
The partial widths are proportional to the strength of the effective $ggH$
operator in Higgs production since we assume an onshell Higgs boson in both
cases, and kinematics cancel for identical Higgs masses. We thus use the same
form factors to estimate the enhancement of $gg \rightarrow H$ production rates
relative to the SM.
\begin{figure} \begin{center}
\includegraphics[width=8.5cm]{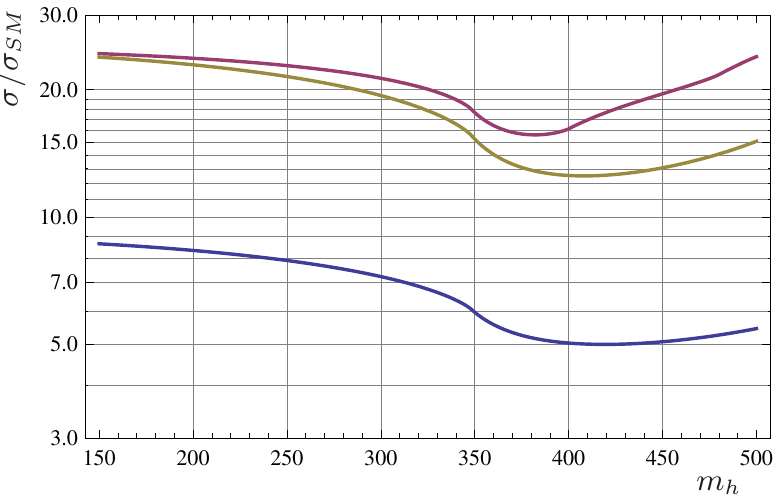}
\end{center}
\caption{\label{fighiggs} The strength of the effective $ggH$ coupling relative
to the SM expressed in terms of production cross sections. The scenarios shown
are (from weakest to strongest) the SM4 ($m_U=310$ GeV, $m_D=260$ GeV), the SM4Q
($m_U=310$ GeV, $m_D=280$ GeV) and again the SM4Q ($m_U=240$ GeV, $m_D=200$
GeV). } 
\end{figure}
The result for three scenarios is shown in Figure \ref{fighiggs}. We observe
that the enhancement is considerably weaker than the naive estimate $\sigma
\propto N_f^2$ which assumes $m_{q'}\rightarrow \infty$. Nevertheless, for most
values of $m_h$, the scenarios with more than one new quark doublet are already
excluded by Higgs searches at the LHC unless one does invoke additional
invisible decay modes for the Higgs which compete with $h\rightarrow WW,ZZ$. For
the scenario without additional quarks, there is no such constraint.





\clearpage

\section{Conclusions}
We have investigated the simplest classes of anomaly free chiral extensions of
the Standard Model, namely a generalization of the usual fourth generation
scenario, a family of two and three quark doublets, and a family of exotic
leptons.  From our renormalization group analysis we conclude that the
experimental lower bounds on fourth generation quark masses have already forced
such models into a regime where additional new physics beyond the extra
generation is necessary below a scale of $10^2\dots 10^3$ TeV. A fourth
generation containing quarks is not compatible with a scenario of grand
unification, unless one invokes a fixed point for strong Yukawa couplings for
which there is no evidence so far. The neutrinoless fourth generation of quarks
faces particularly strong constraints, since the enhanced chiral particle
content (six new doublets rather than four as in the SM4 case) both contributes
very significantly to Higgs production cross sections, and results in a lower
mass window. The fourth generation of leptons is less constrained by direct
searches, which allows us to raise the cutoff close to the putative unification
scale $\Lambda \sim 10^{12}$ GeV.  As the experimental bounds for exotic fermion
masses rise, the RGE evolution of the quartic scalar coupling is dominated by
the contributions from fourth generation Yukawa couplings, and the expected
Higgs boson mass is typically above $200\dots 300$ GeV. This has an important
effect on electroweak precision tests. In order to illustrate this, we have 
analyzed the contributions to oblique corrections in several selected scenarios
which remain perturbative above  $\Lambda_c\geq 10$ TeV. We find that
conventional fourth generation models and a color-neutral fourth generation pass
electroweak precision tests for a range of masses, while scenarios with more
than one additional quark doublet pass electroweak precision tests only for
suitable hypercharge assignments.

\section*{Acknowledgements}
We thank A. Hebecker, K. Matchev, T. Plehn and C. Speckner for useful
discussions. AK would like to thank the University of Florida, Gainesville for
hospitality while parts of this work were finished.
\bibliography{fourthgeneration}

\end{document}